# How Standing Electromagnetic Wave Modulates Electron Current

M.A. Kutlan

Institute for Particle & Nuclear Physics, Budapest, Hungary

kutlanma@gmail.com

Interaction of nonrelativistic electrons with a standing electromagnetic wave is considered. The modulation amplitude of an electron current in the field of a standing or traveling electromagnetic wave is calculated in the quantum approach.

## INTRODUCTION

Elastic scattering of nonrelativistic electrons in the field of a standing electromagnetic field was first considered by Kapitza and Dirac [1]. Their effect in the field of a strong standing electromagnetic wave was investigated by Fedorov [2]. Gaponov and Miller [3,4] studied the modulation and acceleration of nonrelativistic electrons in the field of counterpropagating waves that are shifted in frequency, and it was noted in [3] that the electron interaction with the standing wave is described by an effective potential that is quadratic in the electromagnetic-field strength.

The emission and absorption in an effective potential, however, were not considered in the cited studies. They were first investigated in [5,6], where the probability of this emission was estimated in first-order perturbation theory. It was shown that in this case the emission probability is proportional to $(V_0/E)^2$, where $V_0$ is the effective potential produced by the standing wave, and E is the electron kinetic energy $E > V_0$.

The electron channeling was investigated in intense standing light wave in [7-9]. In [10] was considered quantum modulation of a current of slow electrons reflected from a vacuum-dielectric interface, and also by elastic reflection of electrons from the surface of a transparent single crystal (Bragg reflection). The TCE spectra and intensity were calculated. Other aspects could be found in [11-62].

One can expect the current-density modulation depth and the emission intensity to increase if the electrons are simultaneously acted upon by a spatially periodic field of a diffraction grating and by a traveling electromagnetic wave. The energy and momentum conservation laws are then simultaneously satisfied on account of the grating quasimomentum. The role of this grating can be assumed, for example, by a standing electromagnetic wave.

We consider in the present paper the modulation (classical and quantum) of an electron current in the field of a standing electromagnetic wave.

**BASIC RELATIONS**

Consider a nonrelativistic electron in the field of a standing linearly polarized wave. We define the wave field by a vector potential

$$A_1 = A_{01}^z \sin \omega_1 t \cos k_1 y, \qquad (1)$$

where $A_{01}^z$ is the amplitude of the vector potential, the superscript $z$ indicates the polarization direction, while $\omega_1$ and $k_1$ are respectively the frequency and field vector of waves counterpropagating along the y axis, and their superposition produces the standing wave.

Let the electron momentum be directed along the y axis. In this geometry, the interaction of the electron with the wave is due to the term $(eA_1)^z$, and the Schrodinger equation for the particle in the field of the wave (1) is

$$i\hbar \frac{\partial \Psi}{\partial t} = -\frac{\hbar^2}{2m} \Delta \Psi + V_0 \cos^2 k_1 y \Psi. \qquad (2)$$

Here $V_0 = (eA_{01}^z)^2 / 2mc^2$ is the effective potential and $A_{01}^z = -\varepsilon_1 \lambdabar_1$, where $\varepsilon_1$ is the electromagnetic field and $\lambdabar_1$ is the wavelength.

Note that the high-frequency ($2\omega_1$) term in the interaction is determined by a phase transformation of a Psi function and is disregarded hereafter.

We represent Eq. (2), accurate to an inessential phase shift, in the form

$$i\hbar \frac{\partial \Psi}{\partial t} = -\frac{\hbar^2}{2m} \frac{\partial^2 \Psi}{\partial y^2} + \frac{V_0}{2} \cos qy \Psi, \qquad (3)$$

where $q = 2k_1$. We add to the standing electromagnetic wave a traveling wave with a vector potential

$$A_2 = A_{02}^y \sin(\omega_2 t - k_2 z), \qquad (4)$$

where $A_{02}^y$ is the amplitude of the vector potential, the superscript y indicates the polarization direction, while $\omega_2$ and $k_2$ are respectively the frequency and the wave vector of the traveling wave $A_{02}^y = -\varepsilon_0 c / \omega_2$.

The Schrodinger equation takes now the form

$$i\hbar \frac{\partial \Psi}{\partial t} = -\frac{\hbar^2}{2m}\frac{\partial^2 \Psi}{\partial y^2} + i\frac{e\hbar}{mc}\sin(\omega_2 t - k_2 z)A_{02}^y \frac{\partial \Psi}{\partial y} + \frac{V_0}{2}\cos qy \Psi. \tag{5}$$

[We have neglected in (5) the high-frequency ($2\omega_2$) term determined by a phase transformation of a Psi function and disregarded hereafter, since it contains the small parameter $eA_{02}^y / mc^2$ compared with the retained terms].

We seek the solution of (5) in the form

$$\Psi(y,t) = \sum_n a_n(y)\Psi_n(y)\exp[ink_2 z - iE_n t / \hbar], \tag{6}$$

where $E_n = E_0 + n\hbar\omega_2$.

We seek the wave function $\Psi_n(y)$ in the semiclassical approximation

$$\Psi_n(y) = \exp\left[\frac{i}{\hbar}\int^y (p_n^2 - 2meV_0 \cos qy')^{1/2} dy'\right]. \tag{7}$$

Here $p_n = (p_0^2 + 2mn\hbar\omega)^{1/2}$ and $p_0(2mE_0)^{1/2}$. We assume hereafter $\omega_2 = \omega$, $k_2 = k$.

The function $a_n(y)$ must be determined. We find it for $E \gg V_0$ by using an approximation from [9-10]. In this approximation, first, $(p_n, p) \gg p_n - p \pm q$, a condition met in the case of resonance $p_n - p \pm q = 0$, and second, $p_n \approx p$, which introduces in $a_n(y)$ an error order $n\hbar\omega / E$. Taking this into account, we get

$$\frac{da_n(y)}{dy}\frac{\partial \Psi_n}{\partial y} = -\frac{e\varepsilon_0}{2\hbar\omega}\left(a_{n+1}\frac{\partial \Psi_{n+1}}{\partial y} - a_{n-1}\frac{\partial \Psi_{n-1}}{\partial y}\right). \tag{8}$$

Expanding $p_n$ in terms of the small parameter $V_0 / E_n$ ($p_0 \equiv p, E_0 \equiv E$) and using expression (7) for $\Psi_n$, we get

$$\frac{da_n(y)}{dy} = -\frac{e\varepsilon_0}{2\hbar\omega}\left\{a_{n+1}\exp\left[\frac{i}{\hbar}(p_{n+1}-p_n)y + i\frac{V_0}{4E}\sin qy\right]\right.$$
$$\left. a_{n-1}\exp\left[\frac{i}{\hbar}(p_{n-1}-p_n)y - i\frac{V_0}{4E}\sin qy\right]\right\}. \quad (9)$$

For $V_0/E_n \ll 1$, Eq. (9) takes the simpler form

$$\frac{da_n(y)}{dy} = -\frac{e\varepsilon_0}{8\hbar\omega}\frac{V_0}{E}\left\{a_{n+1}(y)\exp\left[\frac{i}{\hbar}(p_{n+1}-p_n-\hbar q)y\right]\right.$$
$$\left. -a_{n-1}(y)\exp\left[\frac{i}{\hbar}(p_{n-1}-p_n+\hbar q)y\right]\right\}. \quad (10)$$

We have thus obtained for $a_n(y)$ the finite-difference differential equation (10).

**MODULATION OF THE ELECTRON-CURRENT DENSITY**

We shall solve (10) for the case n = + 1. Assuming that $|a_{\pm 1}| \ll 1$ and $a_0 \approx 1$, we obtain, if the interaction is turned-on instantaneously ($Lq \gg 1$)

$$a_{\pm 1}(y) = \mp\frac{e\varepsilon_0}{8\hbar\omega}\frac{V_0}{E}\int_0^y \exp\left[\frac{i}{\hbar}(p-p_{\pm 1}\pm\hbar q)y'\right]dy'$$
$$= \frac{e\varepsilon_0}{4\hbar\omega}\frac{V_0}{E}\exp\left[\frac{i}{\hbar}(p-p_{\pm 1}\pm\hbar q)\frac{y}{2}\right]\times\frac{\sin\left[(p-p_{\pm 1}\pm\hbar q)\frac{y}{2}\right]}{p-p_{\pm 1}\pm\hbar q}. \quad (11)$$

It follows from (11) that $a_1$ increases with y if $p_1 - p = q$ or if $\hbar\omega \ll E$, $\omega/v = q + \pi/L$, where v = p/m is the electron velocity and $L = (8\pi\hbar p/p)(E/\hbar\omega)^2$ is the modulation length [7]' The analogous condition for the increase of the amplitude $a_{-1}$ is $\omega/v = q - \pi/L$. The conditions for $a_1$ and $a_{-1}$ to increase are thus incompatible. That is to say, if the electron beam were ideally monochromatic it would be possible to "tune" it only to absorption ($a_1 \neq 0$, $a_{-1} = 0$) or only to stimulated emission ($a_1 = 0$, $a_{-1} \neq 0$), depending on the relations between the parameters $\omega$, v, and q. This asymmetry of the coefficients is a quantum effect manifested only when the length l of the intersection of the standing wave with the electron beam is much larger than L, since L contains the Planck constant $\hbar$. Let us calculate the electron-current density, using expressions (6) and (7) with $n = \pm 1$ and taking (11) into account:

$$j = j_0 \left(1 - \frac{e\varepsilon_0}{2\hbar\omega} \frac{V_0}{E} \left(\frac{\sin D_{+1} y/2}{D_{+1}} \cos(\varphi - D_{+1} y/2) - \frac{\sin D_{-1} y/2}{D_{-1}} \cos(\varphi - D_{-1} y/2)\right)\right), \quad (12)$$

where $j_0$ is the incident-beam current density, defined by

$$j_0 = \frac{I_0}{\pi ab} \exp\left(-\frac{x^2}{a^2} - \frac{z^2}{b^2}\right).$$

Here $I_0$ is the total beam current, and a and b are the effective beam widths in the *x* and z directions,

$$D_{\pm 1} = \frac{\omega}{v} - q \mp \frac{\pi}{L}, \quad \varphi = \omega t - qy - kz - \frac{V_0}{2E} \sin qy.$$

A real electron beam, however, is not monochromatic, so that expression (12) for the current density must be averaged over the initial-electron-beam distribution function *F(v)* in the velocities. We put

$$F(v) = \frac{1}{\sqrt{\pi}\Delta v} \exp\left(-\frac{(v - v_0)^2}{(\Delta v)^2}\right),$$

where $\Delta v$ is the electron-velocity scatter and $v_0$ is the average electron veldcity in the beam ($\Delta v \ll v_0$). The averaged current can be written in the form

$$j = j_0 \left[1 - \frac{e\varepsilon_0}{4\hbar\omega} \frac{V_0}{E} \operatorname{Re}\left\{e^{i\varphi}(f_{+1}(y) - f_{-1}(y))\right\}\right] \quad (13)$$

where

$$f_{\pm 1} = 2\int_{-\infty}^{+\infty} F(v) \frac{\sin D_{\pm 1} y/2}{D_{\pm 1}} \exp(-iD_{\pm 1} x/2) dv = \int_0^{y} \int_{-\infty}^{+\infty} F(v) \exp(-iD_{\pm 1} y') dy' dx$$

**References**

1. P. L. Kapitza and P. A. M. Dirac, Proc. Cambr. Phil. Soc. 29, 297 (1933).
2. M. V. Fedorov, Zh. Eksp. Teor. Fiz. 52,1434 ( 1967).
3. A. V. Gaponov and M. A. Miller, Zh. Eksp. Teor. Fiz. 34,242 (1958)

4. A. V. Gaponov and M. A. Miller, *ibid.* 751 [515].
5. Fedorov M.V., Oganesyan K.B., Prokhorov A.M., Appl. Phys. Lett., **53**, 353 (1988).
6. Oganesyan K.B., Prokhorov A.M., Fedorov M.V., Sov. Phys. JETP, **68,** 1342 (1988).
7. Oganesyan KB, Prokhorov AM, Fedorov MV, Zh. Eksp. Teor. Fiz., 53, 80 (1988).
8. D. F. Zaretskii and Yu. A. Malov, Zh. Eksp. Teor. Fiz. 91, 1302 (1986).
9. D. F. Zaretskii, Yu. A. Malov, Zh. Eksp. Teor. Fiz. **97**,162 (1990).
10. D. F. Zaretskii, Yu. A. Malov, Zh. Tekh. Fiz. 56. 1256 (1986.)
11. Petrosyan M.L., Gabrielyan L.A., Nazaryan Yu.R., Tovmasyan G.Kh., Oganesyan K.B., Laser Physics, **17**, 1077 (2007).
12. A.H. Gevorgyan, M.Z. Harutyunyan, K.B. Oganesyan, E.A. Ayryan, M.S. Rafayelyan, Michal Hnatic, Yuri V. Rostovtsev, G. Kurizki, arXiv:1704.03259 (2017).
13. K.B. Oganesyan, J. Mod. Optics, **61,** 763 (2014).
14. L.A.Gabrielyan, Y.A.Garibyan, Y.R.Nazaryan, K.B. Oganesyan, M.A.Oganesyan, M.L.Petrosyan, A.H. Gevorgyan, E.A. Ayryan, Yu.V. Rostovtsev, arXiv:1704.004730 (2017).
15. D.N. Klochkov, AI Artemiev, KB Oganesyan, YV Rostovtsev, MO Scully, CK Hu, Physica Scripta, **T140,** 014049 (2010).
16. AH Gevorgyan, MZ Harutyunyan, KB Oganesyan, MS Rafayelyan, Optik-International Journal for Light and Electron, Optics, 123, 2076 (2012).
17. D.N. Klochkov, AI Artemiev, KB Oganesyan, YV Rostovtsev, CK Hu, J. of Modern Optics, **57,** 2060 (2010).
18. K.B. Oganesyan, J. Mod. Optics, **62,** 933 (2015).
19. K.B. Oganesyan, M.L. Petrosyan, YerPHI-475(18) – 81, Yerevan, (1981).
20. D.N. Klochkov, A.H. Gevorgyan, K.B. Oganesyan**,** N.S. Ananikian, N.Sh. Izmailian, Yu. V. Rostovtsev, G. Kurizki, arXiv:1704.006790 (2017).
21. K.B. Oganesyan, J. Contemp. Phys. (Armenian Academy of Sciences), **52,** 91 (2017).
22. AS Gevorkyan, AA Gevorkyan, KB Oganesyan, Physics of Atomic Nuclei, **73**, 320 (2010).
23. AH Gevorgyan, KB Oganesyan, EM Harutyunyan, SO Arutyunyan, Optics Communications, **283**, 3707 (2010).
24. E.A. Nersesov, K.B. Oganesyan, M.V. Fedorov, Zhurnal Tekhnicheskoi Fiziki, **56**, 2402 (1986).
25. A.H. Gevorgyan, K.B. Oganesyan, Optics and Spectroscopy, **110**, 952 (2011).
26. AH Gevorgyan, KB Oganesyan, GA Vardanyan, GK Matinyan, Laser Physics, **24,** 115801 (2014)

27. K.B. Oganesyan, J. Contemp. Phys. (Armenian Academy of Sciences), **51,** 307 (2016).
28. AH Gevorgyan, KB Oganesyan, Laser Physics Letters **12** (12), 125805 (2015).
29. DN Klochkov, KB Oganesyan, EA Ayryan, NS Izmailian, Journal of Modern Optics **63,** 653 (2016).
30. K.B. Oganesyan. Laser Physics Letters, **13**, 056001 (2016).
31. DN Klochkov, KB Oganesyan, YV Rostovtsev, G Kurizki, Laser Physics Letters **11,** 125001 (2014).
32. K.B. Oganesyan, Nucl. Instrum. Methods A **812,** 33 (2016).
33. AS Gevorkyan, AA Gevorkyan, KB Oganesyan, GO Sargsyan, Physica Scripta, **T140,** 014045 (2010).
34. AH Gevorgyan, KB Oganesyan, Journal of Contemporary Physics (Armenian Academy of Sciences) **45,** 209 (2010).
35. K.B. Oganesyan. Laser Physics Letters, **12**, 116002 (2015).
36. GA Amatuni, AS Gevorkyan, AA Hakobyan, KB Oganesyan, et al, Laser Physics, **18,** 608 (2008).
37. K.B. Oganesyan, J. Mod. Optics, **62,** 933 (2015).
38. A.H. Gevorgyan, K.B.Oganesyan, E.M.Harutyunyan, S.O.Harutyunyan, Modern Phys. Lett. B, **25**, 1511 (2011).
39. A.H. Gevorgyan**,** M.Z. Harutyunyan, G.K. Matinyan, K.B. Oganesyan, Yu.V. Rostovtsev, G. Kurizki and M.O. Scully**,** Laser Physics Lett., **13,** 046002 (2016).
40. K.B. Oganesyan, J. Mod. Optics, **61,** 1398 (2014).
41. ZS Gevorkian, KB Oganesyan, Laser Physics Letters **13**, 116002 (2016).
42. AI Artem'ev, DN Klochkov, K Oganesyan, YV Rostovtsev, MV Fedorov, Laser Physics **17**, 1213 (2007).
43. A.I. Artemyev, M.V. Fedorov, A.S. Gevorkyan, N.Sh. Izmailyan, R.V. Karapetyan, A.A. Akopyan, K.B. Oganesyan, Yu.V. Rostovtsev, M.O. Scully, G. Kuritzki, J. Mod. Optics, **56**, 2148 (2009).
44. A.S. Gevorkyan, K.B. Oganesyan, Y.V. Rostovtsev, G. Kurizki, Laser Physics Lett., **12**, 076002 (2015).
45. K.B. Oganesyan, J. Contemp. Phys. (Armenian Academy of Sciences), **50,** 312 (2015).
46. M.V. Fedorov, E.A. Nersesov, K.B. Oganesyan, Sov. Phys. JTP, **31,** 1437 (1986).
47. K.B. Oganesyan, M.V. Fedorov, *Zhurnal Tekhnicheskoi Fiziki***,** **57**, 2105 (1987).
48. Zaretsky, D.F., Nersesov, E.A., Oganesyan, K.B., and Fedorov, M.V., Sov. J. Quantum Electronics, **16**, 448 (1986).
49. K.B. Oganesyan, J. Contemp. Phys. (Armenian Academy of Sciences), **50,** 123 (2015).

50. DN Klochkov, AH Gevorgyan, NSh Izmailian, KB Oganesyan, J. Contemp. Phys., **51,** 237 (2016).

51. K.B. Oganesyan, M.L. Petrosyan, M.V. Fedorov, A.I. Artemiev, Y.V. Rostovtsev, M.O. Scully, G. Kurizki, C.-K. Hu, Physica Scripta, **T140**, 014058 (2010).

52. V.V. Arutyunyan, N. Sh. Izmailyan, K.B. Oganesyan, K.G. Petrosyan and Cin-Kun Hu, Laser Physics, **17**, 1073 (2007).

53. E.A. Ayryan, M. Hnatic, K.G. Petrosyan, A.H. Gevorgyan, N.Sh. Izmailian, K.B. Oganesyan, arXiv: 1701.07637 (2017).

54. A.H. Gevorgyan, K.B. Oganesyan, E.A. Ayryan, M. Hnatic, J.Busa, E. Aliyev, A.M. Khvedelidze, Yu.V. Rostovtsev, G. Kurizki, arXiv:1703.03715 (2017).

55. A.H. Gevorgyan, K.B. Oganesyan, E.A. Ayryan, Michal Hnatic, Yuri V. Rostovtsev, arXiv:1704.01499 (2017).

56. Oganesyan K.B., Prokhorov, A.M., Fedorov, M.V., ZhETF, **94**, 80 (1988).

57. E.M. Sarkisyan, KG Petrosyan, KB Oganesyan, AA Hakobyan, VA Saakyan, Laser Physics, **18,** 621 (2008).

58. E.M. Sarkisyan, AA Akopyan, KG Petrosyan, KB Oganesyan, VA Saakyan, Laser Physics, **19,** 881 (2009).

59. M.V. Fedorov, K.B. Oganesyan, IEEE J. Quant. Electr, **QE-21**, 1059 (1985).

60. D.F. Zaretsky, E.A. Nersesov, K.B. Oganesyan, M.V. Fedorov, Kvantovaya Elektron. **13** 685 (1986).

61. A.H. Gevorgyan, K.B. Oganesyan, R.V. Karapetyan, M.S. Rafaelyan, Laser Physics Letters, **10**, 125802 (2013).

62. K.B. Oganesyan, Journal of Contemporary Physics (Armenian Academy of Sciences) **51,** 10 (2016).